\documentclass[a4paper]{jpconf}
\usepackage{graphicx}
\begin{document}
\title{Local shower age and segmented slope parameters
of lateral density distributions of cosmic ray shower
particles}

\author{Rajat K. Dey}

\address{Department of Physics, University of North Bengal, Siliguri, West Bengal, INDIA 734 013}

\ead{rkdey2007phy@rediffmail.com}

\begin{abstract}
The paper describes lateral density distributions of electrons and muons of cosmic ray extensive air showers (EAS) in the energy regime of the KASCADE experiment. Potential EAS observables are extracted while some suitable lateral distribution functions are introduced to describe these lateral density distributions of simulated/observed cosmic ray showers. Mass-sensitivity of these parameters are demonstrated particularly for the KASCADE electron and muon distributions data. Our studies indicate that the KASCADE data favours an idea of gradual change of cosmic ray mass composition
from lighter concentration to heavier concentration in terms of these parameters.
\end{abstract}

\section{Introduction}
Recent studies using EAS data reveal that the lateral distribution (LD) of EAS electrons could not be analyzed accurately with a single age at all radial distances, which hint that the lateral age varies with the radial distance. At this situation, the idea of variable shower age or local shower age parameter (LAP; ${\rm{s}_{\rm{local}}(\rm{r}})$) of an EAS was introduced [1]. Recent studies reported that the nature of variation of the LAP with $\rm{r}$ does not change much with CR energies. This indicates that the radial variation of the LAP exhibits some sort of scaling nature. The slope parameter of the LD of muons also varies with $\rm{r}$. Due to that, the idea of segmented slope parameter (SSP) has been introduced. Here, the SSP has been estimated from the simulation and the KCDC muon data.

\section{Local age and segmented slope parameters}
The analytical expression of the LAP and SSP using some suitable lateral density function (LDF) between two adjacent points [$\rm{x_{i}},\rm{x_{j}}$] is~: 

\begin{equation}
a_{\rm{local}}(i,j) = {{\ln(F_{ij} X_{ij}^{\alpha_{1}} Y_{ij}^{\alpha_{2}})} \over {\ln(X_{ij} Y_{ij})}}
\end{equation}
 
Where, $F_{ij}$ = {{$f(r_{i}$)}/{$f(r_{j}$)}}, $X_{ij}$=$r_{i}$/$r_{j}$, and $Y_{ij}$=($x_{i}$+1)/($x_{j}$+1). If $r_{i} \rightarrow r_{j}$,  the parameter $a_{local}(x)$ (or $a_{local}(r)$) at each point would appear as~:

\begin{equation}
a_{\rm{local}}(x) = {1 \over {2x+1}} \left( (x+1) {{\partial{\ln f}} \over {\partial{\ln x}}} + (\alpha_{1}+\alpha_{2})x + \alpha_{1} \right)
\end{equation}

The identification $a_{local}(r)\equiv a_{local}(i,j)$ for $r=\frac{r_{i} +r_{j}}{2}$ remains valid for the experimental distributions (taking $F_{ij}~ =~\rho(r_{i})/\rho(r_{j}$)$\equiv ~f(r_{i})/f(r_{j}$)).
For estimating LAP for LDs of electrons and muons, the well-known NKG type LDF is exploited. The Greisen type LDF has been used for estimating the SSP from LDs of muons.

For the NKG-type LDF, the exponents $\alpha_{1}$ and $\alpha_{2}$ in the relation (1) take values as $2$ and $4.5$ while for Greisen-type LDF, $\alpha_{1}$ and $\alpha_{2}$ would take $0$ and $2.5$ respectively. Moreover, we have modified the denominator in the relation (1) by $X_{ji} = X_{ij}^{-1}$ and $Y_{ij} = 1$ while Greisen-type LDF is being considered. 

 For the NKG-type LDF, the $a_{\rm{local}}(r)$ offers the so called  LAP, is already denoted by $s_{\rm{local}}(r)$. Similarly for the Greisen-type LDF i.e. $f_{\rm{Greisen}}(r)$, we denote the shape parameter by $\beta_{\rm{ss}}(r)$, and is called the SSP.

\section{Analysis of Simulation and KCDC data}

In simulation the EPOS 1.99 interaction model for the high-energy hadronic interactions, in combination with the UrQMD model for the low-energy hadronic interactions, embedded in CORSIKA ver. 7.400 were used [2]. A smaller sample of events has also been generated with combinations, QGSJet01.c - UrQMD and EPOS-LHC -UrQMD. The EGS4  program package has taken  care of the EM interactions in the EAS.

We have computed the LAP for the LDs of electrons and muons for each shower separately by using equation (1). 
The KASCADE data from KCDC [3] are available in the form of deposited electron and muon energies in the unshielded and shielded scintillation detectors. We have obtained these data from KCDC, distributed in various detector positions, from near $10$ m to far $200$ m and beyond. The muon LD data in the KASCADE experiment were fitted in the $40 - 200$ m radial distance range to obtain EAS parameters. These fitted parameters are available in a different root file in KCDC. It is noticed that the quality of muon data available in KCDC root file are less precise beyond $\approx{120}$ m range for estimating the LAP and SSP systematically. The main observables like the mean minimum LAP and the mean maximum SSP, have been effectively estimated at about 44 m (LAP) and 71 m (SSP) respectively. The deposited energy data obtained from detectors have been converted to electron and muon numbers and then to their densities in the analysis. For converting energy data into densities, we have used their so-called lateral energy correction functions for electrons and muons available in the KASCADE report. 

\section{Results and discussions}
The variation of the LAP and SSP with radial distance from the EAS core is a fundamental study here. Fist time, the LD of muons has been analyzed for estimating LAP/SSP, thereby exploring the primary mass sensitivity of these parameters with the KASCADE muon data. 

The variation of $s_{\rm{local}}(r)$ with $r$ is shown in Fig. 1a and 1b for electron size intervals. The figure reveals that the experimental data clearly support the trend predicted by the simulation results at intermediate radial distances. We study the above radial variation of the LAP for simulated and KASCADE muon LD data, which is shown in Fig. 1c and 1d. Interestingly, the shapes of the radial distance versus local age curves for muons estimated from the simulated LD data exhibit almost the same configuration. Whatever may be the high-energy hadronic interaction models employed, it is noticed that the LAP estimated from the KASCADE muon data could not follow the simulated results exactly.  Though all figures are not shown, but we have noticed that the radial variation of the local age follows a nearly high-low-high-low type of common feature for both the species, electrons and muons.

On the other hand, in Fig. 2, such type of study is depicted for the parameter $\beta_{\rm{ss}}(r)$ for muon LDs. Here, it appears that the KASCADE results follow the simulation predictions convincingly from both the high energy hadronic models. In these figures, we have noticed an opposite trend in the variation of the $\beta_{\rm{ss}}(r)$ parameter relative to $s_{\rm{local}}(r)$ versus $r$. It can be concluded that the LAP and the SSP of electrons and muons exhibit some sort of scaling behavior while varied as a function of the core distance.  

\begin{figure}[h]
\begin{minipage}{18pc}
\includegraphics[width=18pc]{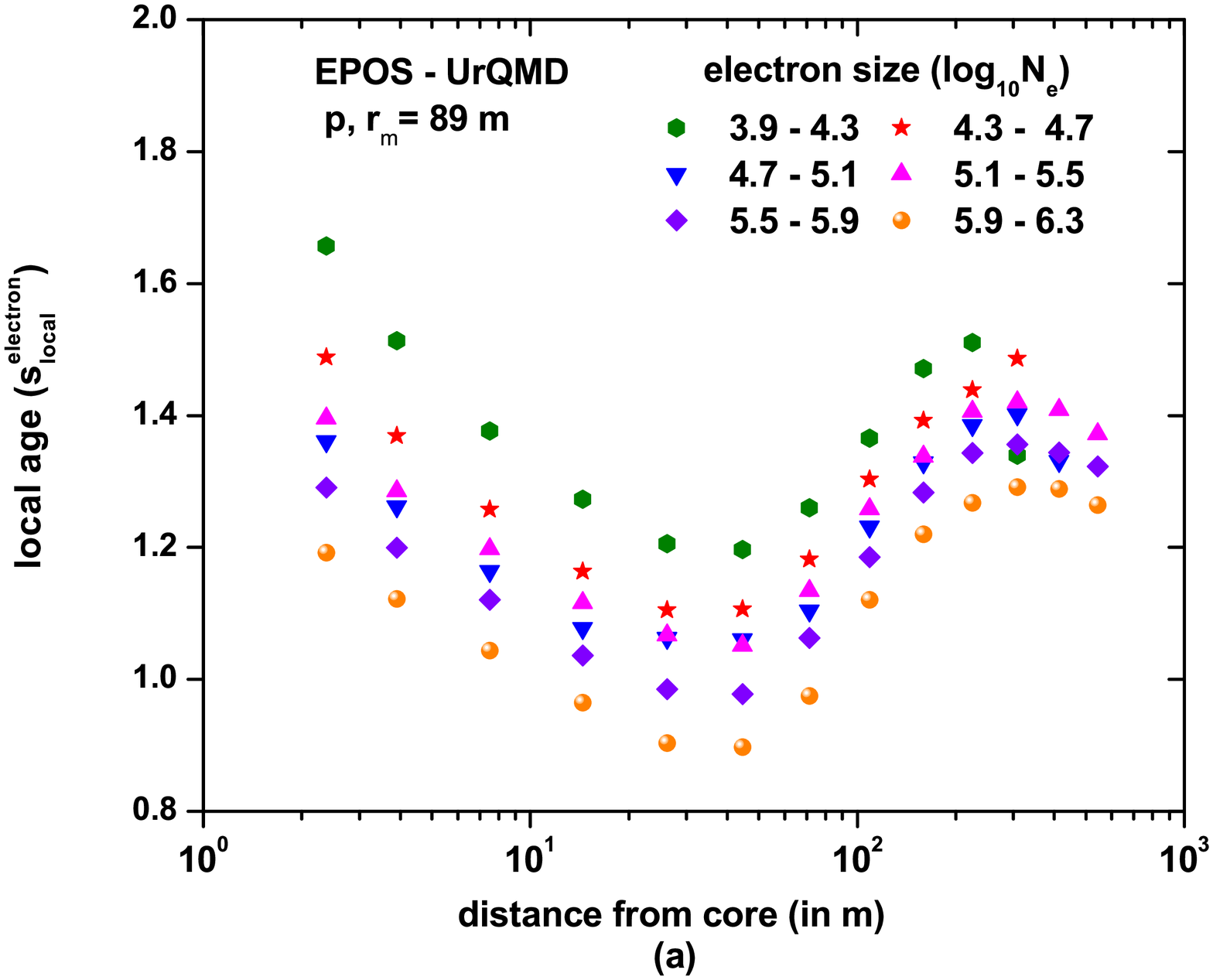}
\end{minipage}\hspace{1pc}%
\begin{minipage}{19pc}
\includegraphics[width=19pc]{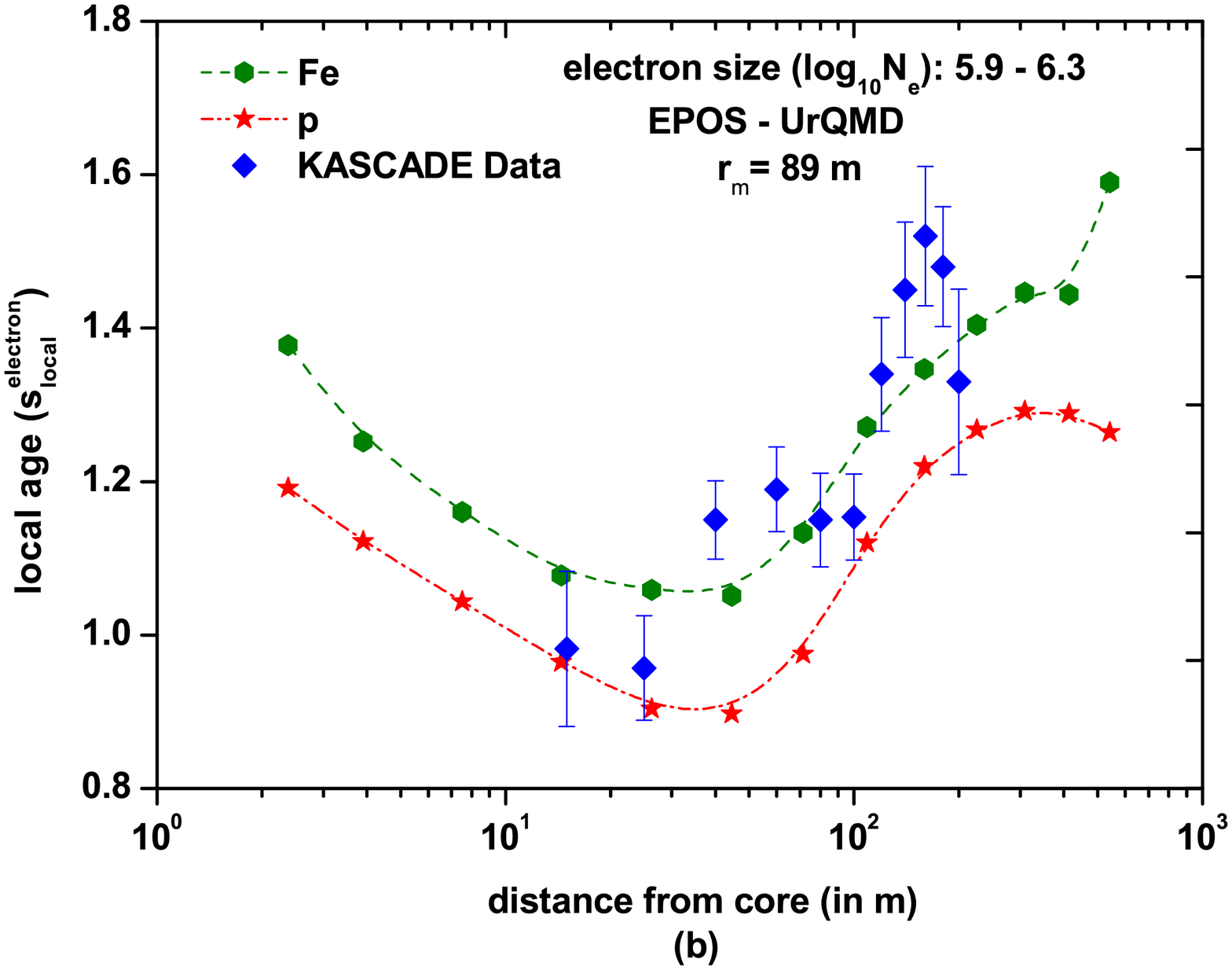}
\end{minipage}\hspace{1pc}%
\begin{minipage}{18pc}
\includegraphics[width=18pc]{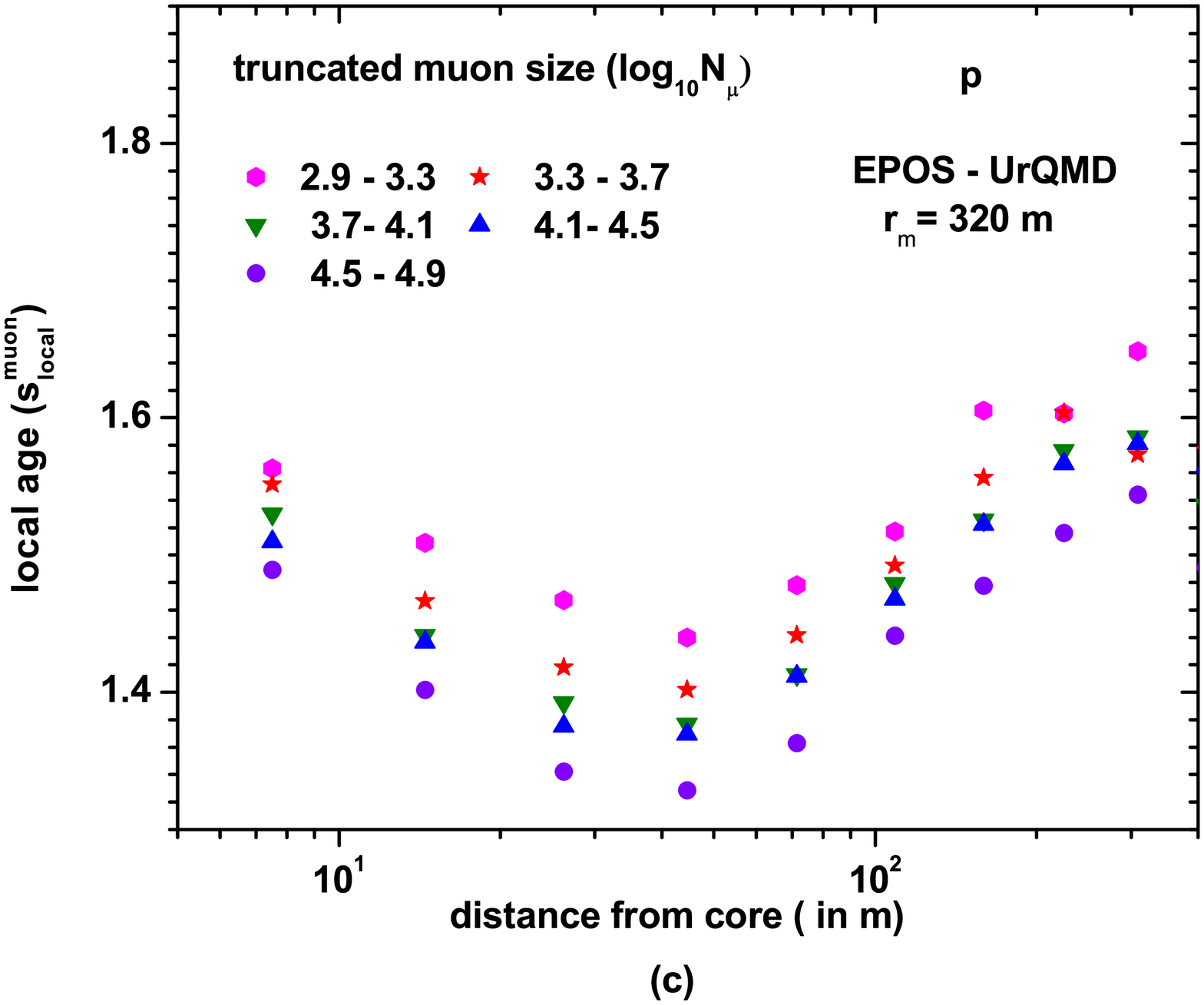}
\end{minipage}\hspace{1pc}%
\begin{minipage}{18pc}
\includegraphics[width=18pc]{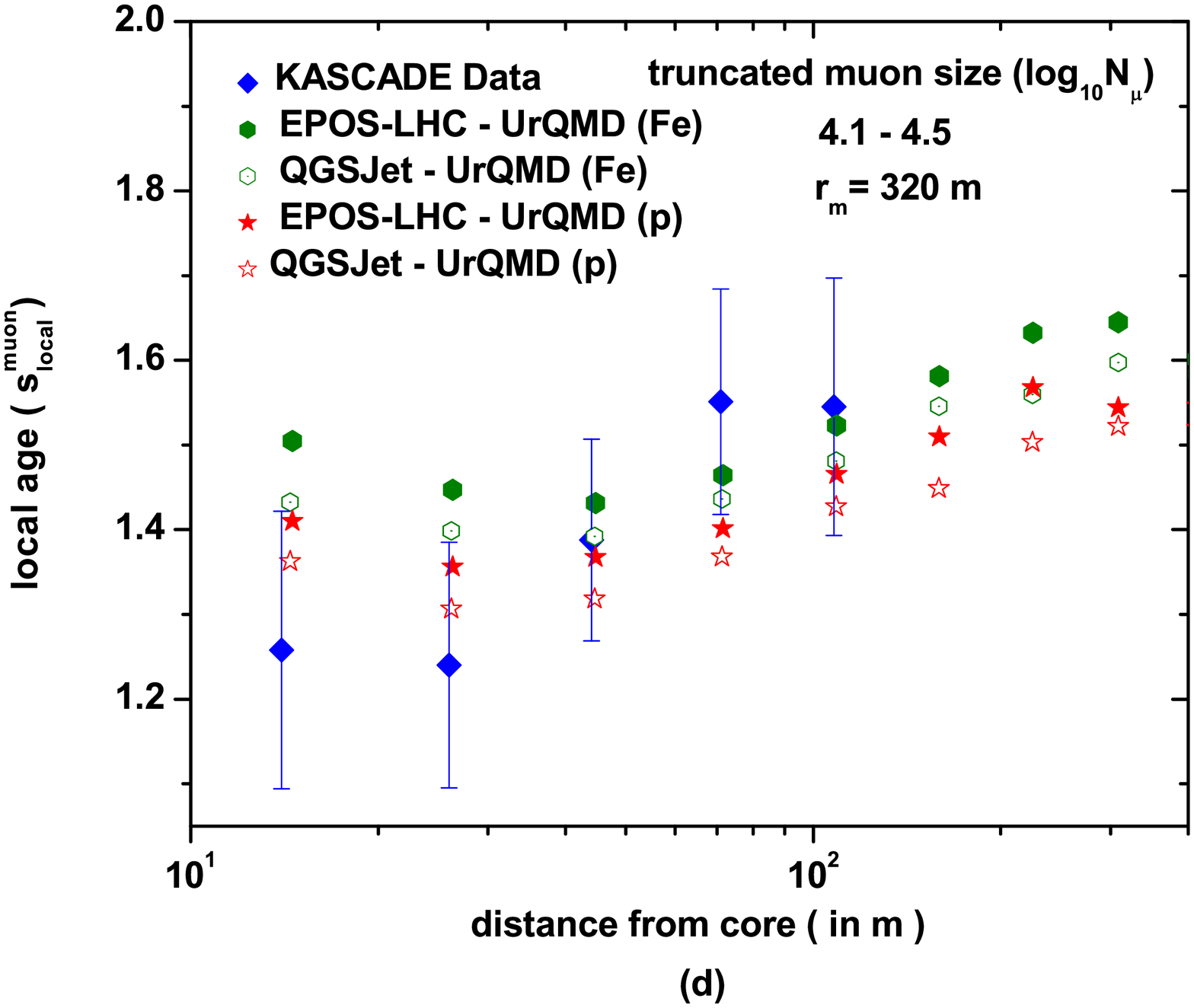}
\caption{\label{label}LAP versus radial distance plot for a mean $N_{\rm e}$ and $N_{\mu}^{tr.}$ interval. In Fig. a and c simulation only while in Fig. b and d simulation and data.}
\end{minipage}\hspace{1pc}%
\end{figure}

\begin{figure}[h]
\begin{minipage}{18pc}
\includegraphics[width=18pc]{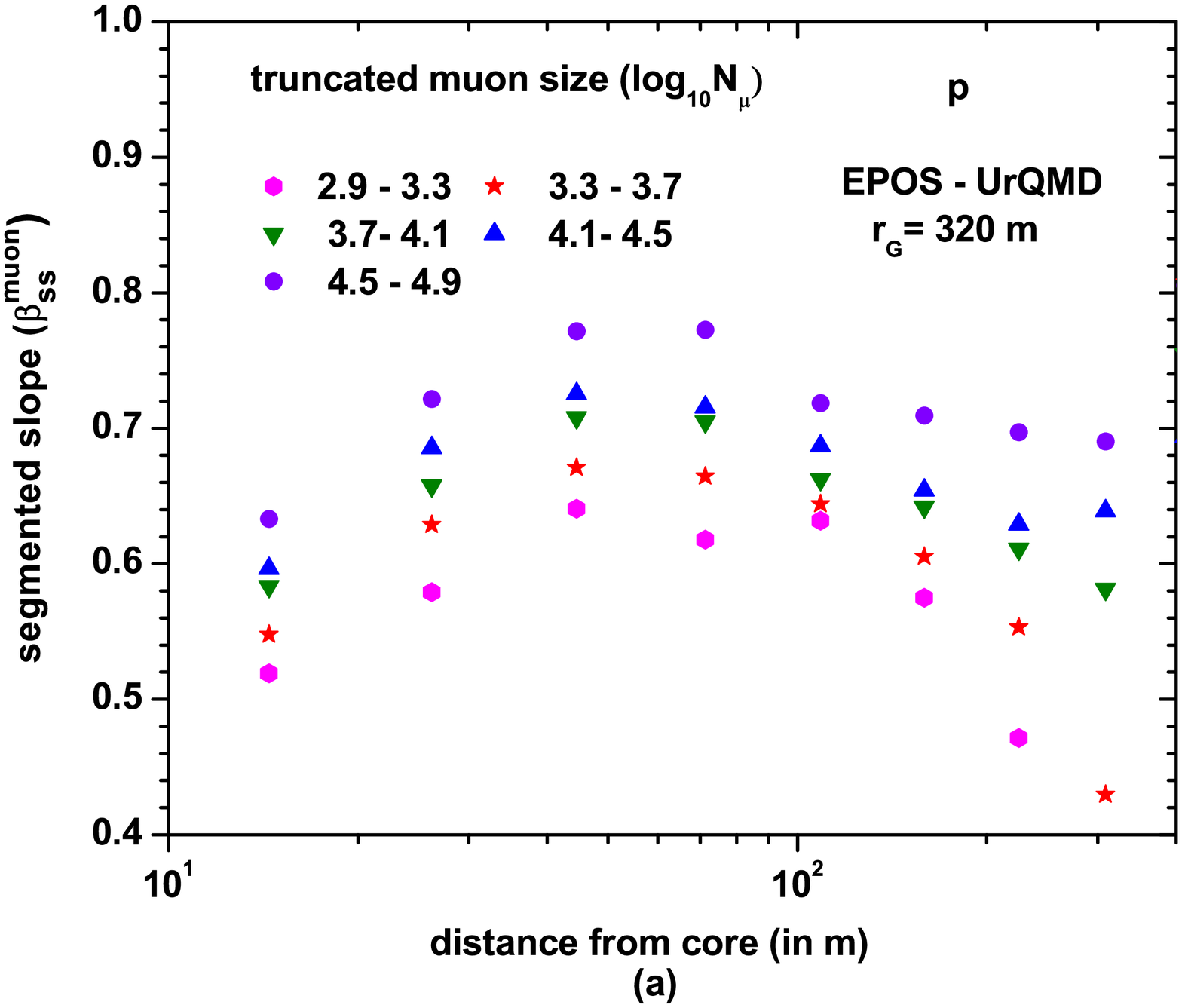}
\end{minipage}\hspace{1pc}%
\begin{minipage}{18pc}
\includegraphics[width=18pc]{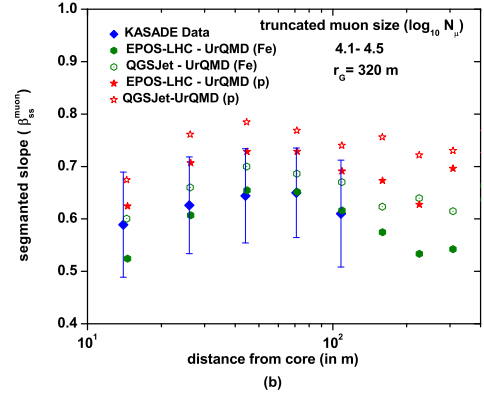}
\caption{\label{label}SSP versus radial distance plot for a mean $N_{\mu}^{tr.}$ interval. In Fig. a for simulation only while Fig. b for simulation and data.}
\end{minipage}\hspace{1pc}%
\end{figure}

\begin{figure}[h]
\begin{minipage}{12pc}
\includegraphics[width=12pc]{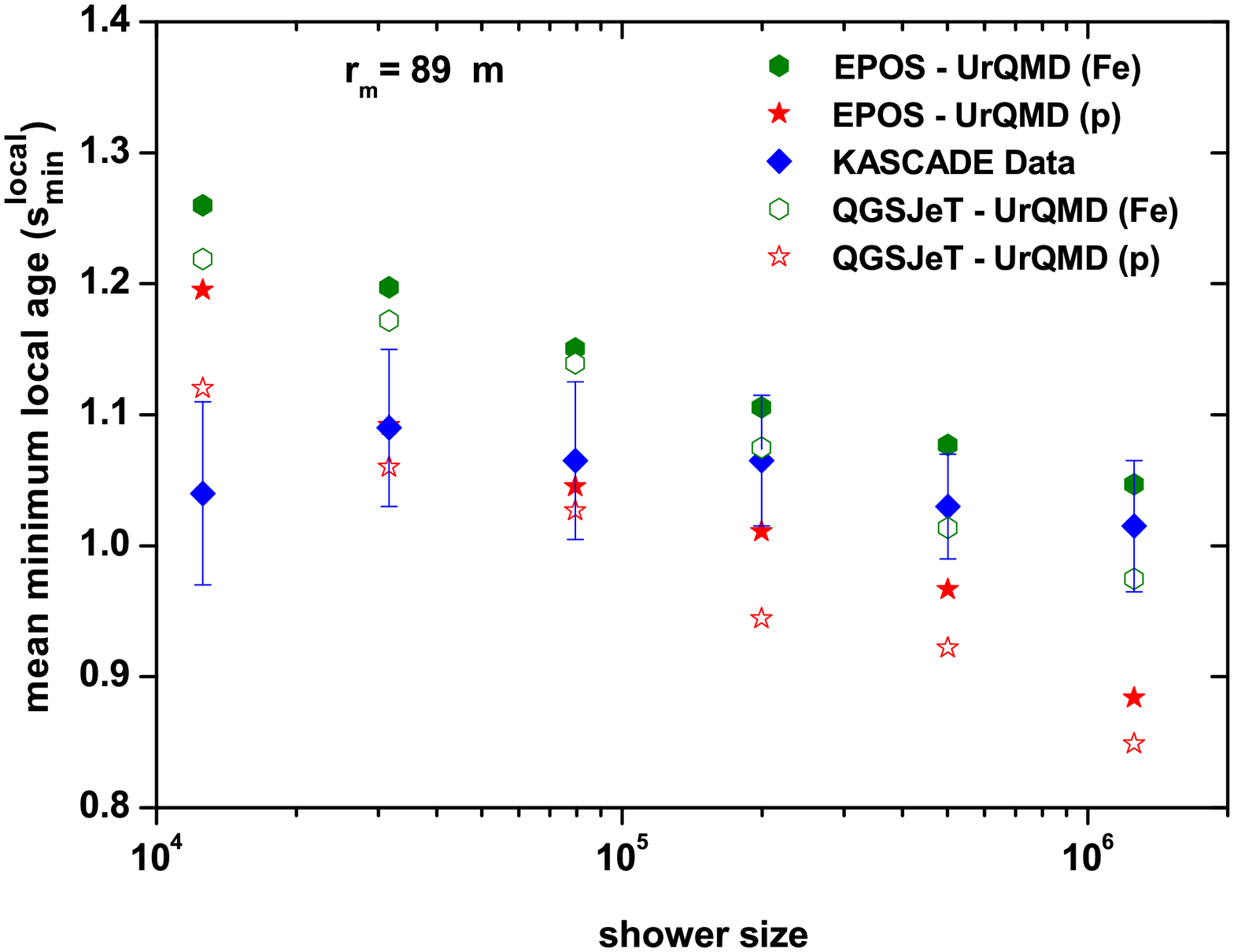}
\caption{\label{label}Mean minimum LAP versus mean $N_{\rm e}$ plot from electron LD data.}
\end{minipage}\hspace{1pc}%
\begin{minipage}{12pc}
\includegraphics[width=12pc]{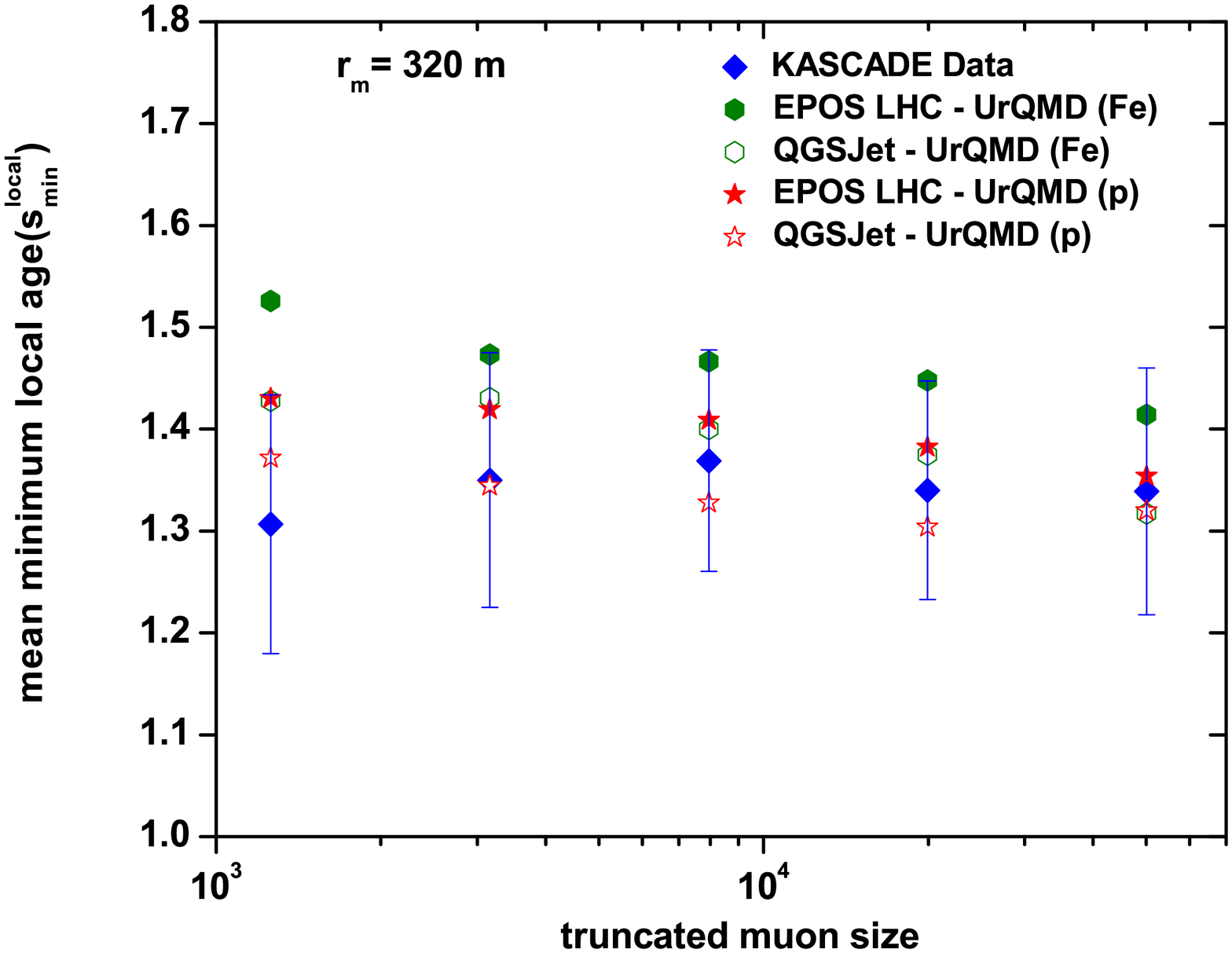}
\caption{\label{label}Mean minimum LAP versus mean $N_{\mu}^{tr.}$ plot from muon LD data.}
\end{minipage}\hspace{1pc}%
\begin{minipage}{12pc}
\includegraphics[width=12pc]{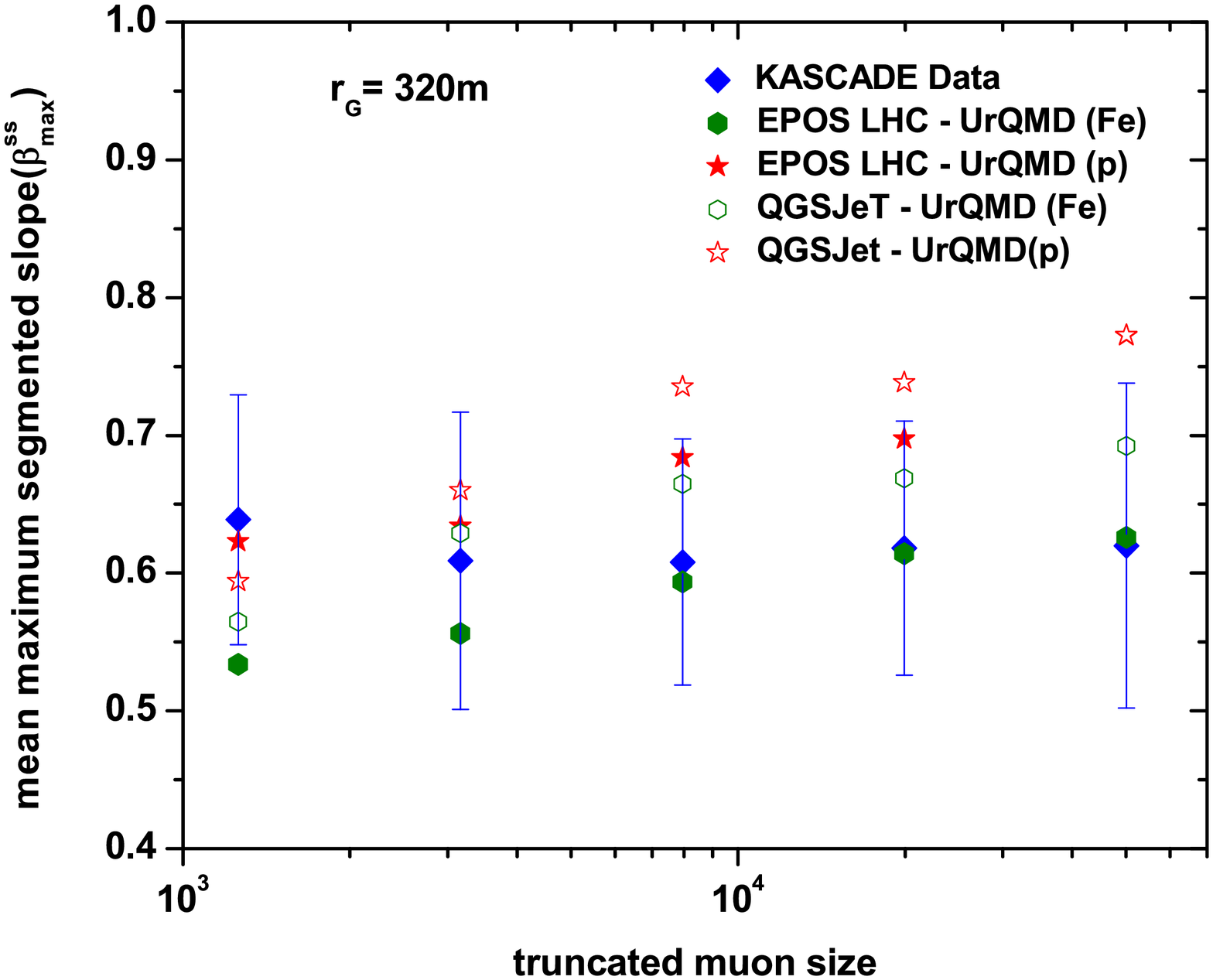}
\caption{\label{label}Mean maximum SSP versus mean $N_{\mu}^{tr.}$ plot from muon LD data.}
\end{minipage} 
\end{figure}

Now, the LAP and SSP versus $r$ curves have been used to explore the CR mass sensitivity of these parameters. For this purpose, we have judiciously assigned a single LAP and SSP to each shower by taking averages of several LAP and SSP at around $44$ m and $71$ m respectively. We label them as mean minimum LAP and mean maximum SSP. These two parameters appear very robust as far as their mass sensitivities are concerned. We have shown the mean minimum local age versus mean shower size plot for electron LD data in Fig. 3. In Fig. 4 such a variation of the mean minimum local age with truncated muon size is also shown for muon LD data. The variation of the mean maximum   
SSP with truncated muon size is shown finally in Fig. 5.
\section{Conclusions}
The LDs of electrons and muons of an EAS manifest some scaling nature in terms of the LAP and SSP. The radial variation of the LAP follows a configuration where with an increasing of the core distance, the parameter decreases initially and attains a minimum, at about $44$ m, then it moves up, attaining a local maximum at about $\approx{300}$ m, and then moves towards a minimum value again. This feature does not change considerably with the shower size and muon size of the EAS. These characteristic variations in the LAP appear as a basic feature of the EM and muonic cascades of a shower. The characteristic feature of SSP versus the core distance curves satisfy a complete opposite configuration compared to the radial variation of the LAP. The SSP estimated from the LDD of muons follows a low-high-low kind of radial variation at least within the range $10 - 300$ m. Here, the SSP attains its maximum value at the core distance $\approx 71$ m. The KASCADE data (Fig. 3, 4 and 5) indicate that the CR mass composition displays a transition from predominantly lighter to heavier nuclei across the knee. This above feature supports the results obtained from the study of the $N_{\rm e}\textendash N_{\mu}$ variations in the KASCADE experiment.

\section*{Acknowledgments}
\noindent
Support under Grant no. ITS/2019/003348 from SERB, India is gratefully acknowledged.

\end{document}